\documentclass[reprint,twocolumn]{revtex4}
\usepackage{amsfonts}
\usepackage{amssymb}
\usepackage{amsmath}
\usepackage{hyperref}
\usepackage{latexsym}
\usepackage{epsfig}

\begin{document}

\title{On scattering of CMB radiation on wormholes: kinetic SZ-effect}
\author{Alexander A. Kirillov
and Elena P. Savelova 
} \affiliation{Dubna International
University of Nature, Society and Man, Universitetskaya Str. 19,
Dubna, 141980, Russia
}
\date{}

\begin{abstract}
The problem of scattering of CMB radiation on wormholes is
considered. It is shown that a static gas of wormholes does not
perturb the spectrum of CMB. In the first order by $v/c$ the
presence of peculiar velocities gives rise to the dipole
contribution in $\Delta T/T$, which corresponds to the well-known
kinetic Sunyaev-Zel'dovich effect. In next orders there appears a
more complicated dependence of the perturbed CMB spectrum on
peculiar velocities. We also discuss some peculiar features of the
scattering on a single wormhole.
\end{abstract}


\maketitle



\section{Introduction}

As it was shown recently all features of cold dark matter models (CDM) can
be reproduced by the presence of a gas of wormholes \cite{KS11,KS07}. At
very large scales wormholes behave exactly like very heavy particles, while
at smaller subgalactic scales wormholes strongly interact with baryons and
cure the problem of cusps. Moreover, there are some strong theoretical
arguments, which come from lattice quantum gravity, that the topological
structure of our Universe should have fractal properties \cite{S15}.
Therefore, we may claim that up to date wormholes give the best candidate
for dark matter particles. The final choice between different dark matter
candidates requires the direct observation of  effects related to wormholes.

Presumably cosmological wormholes are not very large (otherwise
they would be directly seen on the sky). In the present paper we
consider the scattering of CMB radiation on wormholes and show
that they can be observed by means of the kinematic
Sunyaev-Zel'dovich effect (kSZ) \cite{ZS,ZSa}. KSZ signal is long
used to study peculiar motions of galaxy clusters and groups and
has already long history, e.g., see \cite{KSZ1,KSZ2,KSZ3} and
references therein. It has a universal nature, i.e., it is
produced by any kind of matter which scatters CMB (not only by hot
gas). In this respect, it is rather difficult separate the
contribution of wormholes into kSZ from that of the electron gas
in clusters and groups. Therefore, we think that one has to look
for such an effect in those spots on the sky where the baryonic
matter is absent, e.g., in voids where the number density of
wormholes should have the biggest value  and the leading
contribution will come from wormholes alone. Indeed, if we accept
the fractal topological structure and the gas of wormholes as the
basic DM candidate, then in voids wormholes push out (or replace)
baryons, which explains why there is no galaxies in voids. There
remains also the possibility to study peculiar features of the
scattering of CMB on wormholes.

As it was demonstrated recently in \cite{S15} stable cosmological wormholes
have throat sections in the form of tori. In the present paper we however
restrict to the simplest spherically symmetric wormholes. The spherical
wormholes have a more symmetric form and can be viewed as a torus averaged
over its orientations. Upon averaging out, some peculiar features of
interest will disappear (we present them elsewhere), nevertheless, the basic
kSZ effect remains.

\section{Cross-sections and kSZ-effect}

The scattering of signals on spherical wormholes has been already studied
e.g., \cite{sct,sct2} see also recent papers \cite{KSWS,KSS,KSZ}. The basic
two important features is the generation of a specific interference picture
(upon scattering on a single wormhole) and the generation of a diffuse halo
around any discrete source. Unfortunately, both features are not so good to
use them in the searching for wormholes (the first one gives too weak
signal, while the second feature may have various interpretations).

Consider first the case of a static gas of wormholes, i.e. in the absence of
peculiar motions. The spherical wormhole can be considered as a couple of
conjugated spherical mirrors, when a relict photon falls on one mirror it is
emitted, upon the scattering, from the second (conjugated) mirror. The
cross-section of such a process has been described by us earlier in \cite%
{KSWS} and can be summarized as follows. Let an incident plane wave (a set
of photons) falls on one throat. Then the scattered signal has the two
parts. First part represents the standard diffraction (which corresponds to
the absorption of CMB photons on the throat) and forms a very narrow beam
along the direction of the propagation. This is the so-called scattering
forward which is described by the cross-section
\begin{equation}
\frac{d\sigma _{absor}}{d\Omega }=\sigma _{0}\frac{\left( ka\right) ^{2}}{%
4\pi }\left\vert \frac{2J_{1}\left( ka\sin \chi \right) }{ka\sin \chi }%
\right\vert ^{2},  \label{abs}
\end{equation}%
where $\sigma _{0}=\pi a^{2}$, $a$ is the radius of the throat, $k$ is the
wave vector, and $\chi $ is the angle from the direction of propagation of
the incident photons, and $J_{1}$ is the Bessel function. Together with this
part the second throat emits an omnidirectional isotropic flux with the
cross-section
\begin{equation}
\frac{d\sigma _{emit}}{d\Omega }=\sigma _{0}\frac{1}{4\pi }.  \label{flux}
\end{equation}%
It is easy to check that the total cross-sections coincide
\begin{equation*}
\int \frac{d\sigma _{absor}}{d\Omega }d\Omega =\int \frac{d\sigma _{emit}}{%
d\Omega }d\Omega =\sigma _{0}
\end{equation*}%
which means the conservation law for the number of photons (the
number of absorbed and emitted photons coincides). This is enough
to understand what is going on with CMB in the presence of the gas
of wormholes. For the static gas (in the absence of peculiar
motions) every wormhole throat end absorbs photons as the
absolutely black body, while the second end of the throat
re-radiates them in an isotropic manner (with the same black body
spectrum). It is clear that there will not appear any distortion
of the CMB spectrum at all. In other words, we may say that in the
absence of peculiar motions the distortion of the spectrum does
not occur.

Consider now the presence of peculiar motions. The motion of one
end of the wormhole throat  with respect to CMB causes the angle
dependence of the incident
radiation with the temperature%
\begin{equation*}
T_{1}=\frac{T_{\gamma }}{\sqrt{1-\beta _{1}^{2}}\left( 1+\beta _{1}\cos
\theta _{1}\right) }\simeq T_{\gamma }\left( 1-\beta _{1}\cos \theta
_{1}+...\right)
\end{equation*}%
where $\beta _{1}=V_{1}/c$ is the velocity of the throat end and $\beta
_{1}\cos \theta _{1}=\left( \vec{\beta}_{1}\vec{n}\right) $, $\vec{n}$ is
the direction for incident photons. Therefore, the absorbed radiation has
the spectrum%
\begin{equation*}
\rho \left( T_{1}\right) =\rho \left( T_{\gamma }\right) +\frac{d\rho \left(
T_{\gamma }\right) }{dT}\Delta T_{1}+\frac{1}{2}\frac{d^{2}\rho \left(
T_{\gamma }\right) }{dT^{2}}\Delta T_{1}^{2}+...,
\end{equation*}%
where $\rho \left( T_{\gamma }\right) $ is the standard Planckian spectrum
and $\Delta T_{1}\left( \beta _{1}\cos \theta _{1}\right) =T_{1}-T_{\gamma }$%
. It turns out that in the first order by $\beta _{1}$ such an anisotropy
gives no contribution in the re-radiated photons and does not contribute in
the distortion of the spectrum. Indeed, in the reference system in which the
second end of the throat is at rest we have the isotropic flux (\ref{flux})
and, therefore, integrating over the incident angle $\theta _{1}$ we find $%
\int $ $\Delta T\left( \cos \theta _{1}\right) d\Omega =0$. This means that
in the first order by $\beta _{1}$ the second end of the throat radiates (in
the rest reference system) the same black body radiation with the same
temperature $T_{\gamma }$. In next orders by $\beta _{1}$ there appears a
non-vanishing contribution to the distortion of the spectrum $\rho \left(
T_{1}\right) -\rho \left( T_{\gamma }\right) $. However, in next orders a
more important features will appear, when we consider the actual wormhole
sections in the form of tori. Therefore, we leave next orders for the future
research.

Consider now the re-radiation of the absorbed CMB photons. In the first
order by $\beta _{2}=V_{2}/c$ ($V_{2}$ is the peculiar velocity of the
second end of the wormhole throat) it radiates the black body radiation with
the apparent surface temperature (brightness)%
\begin{equation}
T_{2}\simeq T_{\gamma }\left( 1+\beta _{2}\cos \theta _{2}+...\right)
\label{sbr}
\end{equation}%
where $\beta _{2}\cos \theta _{2}=\left( \vec{\beta}_{2}\vec{m}\right) $ and
$\vec{m}$ is the unit vector pointing out to the observer. This is exactly
the kinematic Sunyaev-Zel'dovich effect.

Consider a collection (cloud) of wormhole throats. To obtain the
net energy-momentum transfer between the CMB radiation and the gas
of wormholes we have to average over the wormhole distribution. On
average CMB photons undergo $\tau _{w}$ scatterings, where $\tau
_{w}$ is the projected cloud optical depth due to the scattering.
If $n(r)$ is the number density of
wormholes measured from the center of the cloud, then $\tau _{w}$ is given by%
\begin{equation*}
\tau _{w}=\pi \overline{a^{2}}\int n(r)d\ell
\end{equation*}%
where the integration is taken along the line of sight and
\begin{equation*}
\overline{a^{2}}=\frac{1}{n}\int a^{2}n(r,a)da.
\end{equation*}%
Here $n(r,a)$ is the number density of wormholes depending on the throat
radius $a$. The optical depth $\tau _{w}$ may be also interpreted as
follows. Let $L$ be the characteristic size of the cloud of wormholes. Then
on the sky it will cover the surface $S\sim L^{2}$, while the portion of
this surface covered by wormhole throats is given by
\begin{equation*}
\tau _{w}=\frac{N\pi \overline{a^{2}}}{L^{2}}=\pi \overline{a^{2}}\overline{n%
}L,
\end{equation*}%
where $N$ is the number of wormhole throats in the cloud and $\overline{n}$
is the mean density. Since all wormhole throats have the surface brightness (%
\ref{sbr}) which is different from that of CMB, the parameter $\tau _{w}$
defines (together with the peculiar velocity of the cloud $\beta _{2}$) the
surface brightness of the cloud itself.

\section{The scattering of CMB on a single wormhole}

In the case of a single wormhole we should account for the two
important features. The first feature is the fact that a stable
cosmological wormhole has the throat section in the form of a
torus \cite{S15}. We point out that under the stable cosmological
wormhole we mean here the wormhole, which does not require the
presence of any form of exotic matter (save baryons), but which is
involved in the cosmological expansion. Stability of such a
wormhole is described by the standard Lifshitz theory and from the
qualitative standpoint it does not differ from the development of
cosmological primordial perturbations. Therefore, if such a
wormhole is sufficiently big, then the simplest way to find it is
to look for the direct imprints on CMB map. Indeed,  by means of
kSZ effect a wormhole should produce a ring on CMB map which has
slightly different from the background temperature. In particular,
it was reported recently in \cite{MNR}, that there are, with
confidence level 99.7 per cent, such ring-type structures in the
observed cosmic microwave background. We hope that such structures
could be
 imprints of cosmological wormholes indeed.

The second important feature is that the scattering forward (i.e.
absorption of CMB photons (\ref{abs})) produces much bigger effect
(since $kR\gg 1$, $ka\gg 1$, where $k$ is the wave-vector, $R$ is
the largest, and $a$ is the smallest radiuses of the torus
respectively). This effect corresponds to the standard diffraction
on the torus-like obstacle. In the approximation $\mu =a/R\ll 1$,
where $a$ is the smallest radius of the torus, we may use the flat
screen approximation.

Let the orientation of the torus (the normal to the torus direction ) be
along the $Oz$ axis, i.e. $m=(0,0,1)$. The cross-section depends on the two
groups of angle variables, i.e. the two unit vectors $n_{0}(\phi _{0},\theta
_{0})$ and $n(\phi ,\theta )$. The vector $n_{0}=(\cos \phi _{0}\sin \theta
_{0},\sin \phi _{0}\sin \theta _{0},\cos \theta _{0}$) points to the
direction of the incident photon (i.e., the wave vector is $k_{0}=\frac{%
\omega }{c}n_{0}$), while the vector $n$ corresponds to the
scattered photons. Then the cross-section is given by
\begin{equation*}
\frac{d\sigma }{d\Omega }=\sigma _{R}\sin ^{2}\theta _{0}\frac{\left(
kR\right) ^{2}}{4\pi }\left( \frac{1+\cos ^{2}\theta }{2}\right) \left\vert
F\right\vert ^{2},
\end{equation*}%
where $\sigma _{R}=\pi R^{2}$, and the function $F$ is
\begin{equation*}
F=\left( 1+\mu \right) ^{2}\frac{2J_{1}\left( \left( 1+\mu \right) x\right)
}{\left( 1+\mu \right) x}-\left( 1-\mu \right) ^{2}\frac{2J_{1}\left( \left(
1-\mu \right) x\right) }{\left( 1-\mu \right) x}
\end{equation*}%
where $x=kR\xi $. We also denote
\begin{equation*}
\xi =\left( \sin ^{2}\theta +\sin ^{2}\theta _{0}-2\sin \theta
\sin \theta _{0}\cos \left( \phi -\phi _{0}\right) \right) ^{1/2}
\end{equation*}%
and $J_{n}(x)$ are the Bessel functions. We also averaged $\sigma
$ over polarizations.
Let us expand the kernel $F$ by the small parameter $\mu \ll 1$ which gives%
\begin{equation*}
F\approx 2\mu \left( x\left( \frac{2J_{1}\left( x\right) }{x}\right)
^{\prime }+2\frac{2J_{1}\left( x\right) }{x}\right) .
\end{equation*}%
Using the property $\left( J_{\nu }(x)/x^{\nu }\right) ^{\prime }=-J_{\nu
+1}(x)/x^{\nu }$ we find
\begin{equation*}
F\approx 2\mu \left( -2J_{2}\left( x\right) +2\frac{2J_{1}\left( x\right) }{x%
}\right)
\end{equation*}%
and from the identity $J_{2}\left( x\right)
=\frac{2}{x}J_{1}(x)-J_{0}(x)$ we get $ F\approx 4\mu J_{0}(x) $
which gives
\begin{equation*}
\frac{d\sigma }{d\Omega }=8\sigma _{R}\frac{\left( ka\right) ^{2}}{4\pi }%
\left( 1-\cos ^{2}\theta _{0}\right) \left( 1+\cos ^{2}\theta
\right) \left\vert J_{0}(kR\xi )\right\vert ^{2} .
\end{equation*}%
Since $kR\gg 1$, the above expression shows the presence of
specific ring-type oscillations in the cross-section. Indeed, if
we consider the normal fall of the incident photons, i.e., $\theta
_{0}=0$, then we find $\frac{d\sigma }{d\Omega } \thicksim (1+\cos
^{2}\theta )J_{0}(kR\theta)$.

\section{Discussions}
Now the basic question arises where we should look for wormholes?
This problem resembles the traditional trial in Russian folklore \
("go there, we do not know where and bring us that we do not know
what"). What we should find out it is more or less fair. When we
will have enough sensitivity to see effects of a single wormhole,
we may observe a specific features of the scattering of CMB on a
wormhole section in the form of a torus (specific rings, etc.). In
this respect preliminary result in \cite{MNR} looks very
optimistic though it requires an independent confirmation. The
more straightforward way is to observe the collective KSZ effect
discussed above. There we meet two basic problems. First one is
the unfair predictions of such an effect. Indeed, we know a little
about the
density of wormholes $n_{w}$ and the characteristic size of sections $%
\overline{a}$ (or $\sigma _{0}=\pi \overline{a^{2}}$). The fractal
distribution of galaxies and the behavior of dark matter in
galaxies may fix the two another parameters \cite{S15} by means of
measuring the empirical Green function
$$
G_{emp}=\frac{-4\pi }{k^{2}( 1+(Rk)^{-\alpha })}
$$
which describes deviation from the Newton's law (at small scales ($Rk\gg 1$)
it gives the standard Newton's law, while at large scales $Rk\ll 1$ it
transforms to the fractal law, or logarithmic behavior). We point out that
the logarithmic correction \ observed in galaxies corresponds to the value $%
\alpha \approx 1$ and $R\sim 5Kpc$ \cite{KT}. However, it is still not quite
clear how we may extract the safe estimate for the optical depth $\tau _{w}$%
. The gas of wormholes may be described by at least tree scale parameters
(the characteristic size of the throat section $\overline{a}$, the number
density $n_{w}$, and the characteristic distance between wormhole entrances $%
\ell $), therefore, the two parameters ($R$ and $\alpha $) are not enough to
fix all parameters of the gas of wormholes and define $\tau _{w}$. In other
words, this problem requires the further study.

The second problem closely relates to the problem where we should
look for wormhole effects. In galaxies and clusters (as well in
the hot X-ray gas) the KSZ effect based on wormholes mixes with
that on other sorts of matter (dust, hot gas, etc.). Therefore,
the best way to look for wormhole effects is to search them in
voids, where the baryonic matter is  less dense.

\end{document}